\documentclass{andp2012}
\usepackage[english]{babel}
\usepackage{cite,bm}
\usepackage{lipsum}

\newcommand{\tr}{\ensuremath{\operatorname{tr}}}

\newcommand{\void}[1]{}
\title{Capacitively coupled nano conductors}
\subtitle{Ratchet currents and exchange fluctuation relations}
\author[R. Hussein]{Robert Hussein}
\author[S. Kohler]{Sigmund Kohler\footnote{Corresponding author\quad
E-mail:~\textsf{sigmund.kohler@icmm.csic.es}}}
\address{Instituto de Ciencia de Materiales de Madrid, CSIC, 28049
Madrid, Spain}
\shortauthors{R. Hussein et al.}
\begin{abstract}
	We investigate electron transport in two quantum circuits with mutual
	Coulomb interaction.  The first circuit is a double quantum dot connected
	to two electron reservoirs, while the second one is a quantum point contact
	in the weak tunneling limit.  The coupling is such that an electron in the
	first circuit enhances the barrier of the point contact and, thus, reduces
	its conductivity.  While such setups are frequently used as charge
	monitors, we focus on two different aspects.  First, we derive transport
	coefficients which have recently been employed for testing generalized
	equilibrium conditions known as exchange fluctuation relations.  These
	formally exact relations allows us to test the consistency of our master
	equation approach.  Second, a biased point contact entails noise on the DQD
	and induces non-equilibrium phenomena such as a ratchet current.
\end{abstract}
\shortabstract
\begin{document}
	\maketitle

	\section{Introduction}
	
	\begin{figure}[b]
		\includegraphics[width=\columnwidth]{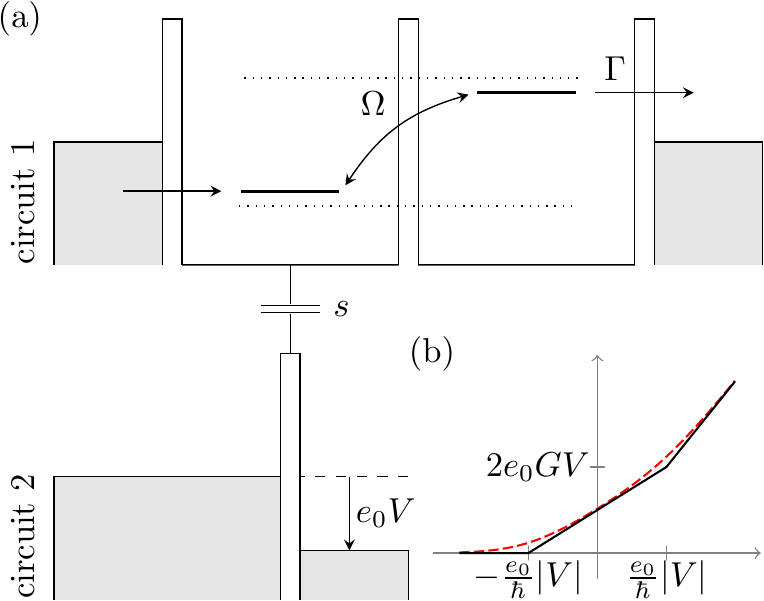}\caption{(a) Double quantum dot (DQD, circuit 1) coupled to a quantum point
		contact (QPC, circuit 2) operated in the weak tunneling regime.  The QPC
		bias voltage $V$ shifts the chemical potential of the right lead by $\mu =
		-e_0V$.
		(b) Correlation function $C(\omega) = C_+(\omega)+C_-(\omega)$ of the
		operator $\Lambda$ by which the QPC couples to the
		left quantum dot for zero temperature (solid line) and $k_BT=0.24eV$
		(dashed), see Eqs.~\eqref{Hqpc} and \eqref{Comega}.}
		\label{fig:setup}
	\end{figure}%
	By now most quantum dots are designed with a nearby quantum point contact
	(QPC) whose conductance is affected by the charge state of the dot
	\cite{Ihn2009a, Taubert2011a}.  Then the QPC can act as monitor for the dot
	charge and provide time-resolved information from which the counting
	statistics of the electrons flowing through the quantum dot
	\cite{Gustavsson2006a, Fricke2007a} and current correlation functions
	\cite{Ubbelohde2012a} can be reconstructed.
	When several quantum dots are strongly tunnel coupled, the wavefunction of
	their electrons becomes delocalized. Then charge detection corresponds to
	the measurement of the electron position and causes decoherence and
	localization.  A quantitative analysis \cite{Hussein2014b} revealed that
	for a double quantum dot (DQD), good measurement correlations can be
	obtained only at the expense of a backaction strong enough to turn coherent
	inter-dot tunneling into classical hopping.
	
	Charge transport in a QPC in the weak tunnel limit consists of uncorrelated
	events in which an electron jumps from one lead to the other
	\cite{Ingold1992a}.  In technical terms, it represents a Poisson process
	whose fluctuations are non-thermal and known as shot noise
	\cite{Blanter2000a}.  When acting upon a quantum system, these fluctuations
	not only cause decoherence, but also excite electrons and
	drive the system out of equilibrium.  In this way, the QPC may play a
	constructive role.  In a asymmetric DQD (upper circuit in
	Fig.~\ref{fig:setup}), shot noise may induce a ratchet current
	\cite{Khrapai2006a}.
	
	Coupled conductors can also be used for testing
	exchange fluctuation relations \cite{Andrieux2006a, Saito2008a,
	Esposito2009a, Campisi2011a} which are generalized equilibrium relations
	based on the assumption that each lead is in a Gibbs state.  Then forward and
	backward rates are related by a Boltzmann factor which provides
	relations between transport coefficients.  A prominent example is the
	Johnson-Nyquist relation between the conductance and the zero-frequency
	limit of the current correlation function \cite{Andrieux2006a, Saito2008a}.
	While exchange fluctuation relations are exact, practical
	computations of transport properties often rely on approximations such as
	perturbation theory in the dot-lead tunneling, which may violate exact
	formal relations. In turn, exact fluctuation relations can be used to test
	the consistency of theoretical methods.  In this spirit, it has been shown
	\cite{Hussein2014a} that the Bloch-Redfield master equation
	\cite{Redfield1957a}---despite being a successful, widely applied, and
	fairly reliable approach---is not always fully compatible with exchange
	fluctuation relations.
	
	In this work, we address two aspects of transport in capacitively coupled
	conductors: First, we compute a set of transport coefficients which can be
	used to experimentally test fluctuations relations along the lines of
	Refs.~\cite{Nakamura2010a, Utsumi2010a}.  Moreover, we investigate to which
	extent our master equation results agree with exact formal relations.
	Second, we characterize the non-equilibrium current through a DQD induced
	by the coupling to a QPC and study cross correlations between these
	subsystems.
	
	\section{Model and master equation approach}
	\label{sec:model}
	
	\subsection{DQD coupled to a QPC in the tunnel limit}
	
	The model sketched in Fig.~\ref{fig:setup} consists of two electric
	circuits, the upper one being a DQD coupled to electron source and drain.
	We describe the DQD by two levels tunnel coupled with matrix element
	$\Omega$ and detuned by $\epsilon$ such that the one-electron states are
	split by $\Delta=\sqrt{\epsilon^2+\Omega^2}$. Thus,
	\begin{equation}
		H_\text{DQD} = \frac{\epsilon}{2}(c_2^\dagger c_2 - c_1^\dagger c_1)
		+ \frac{\Omega}{2}(c_1^\dagger c_2 + c_2^\dagger c_1) + U n_1 n_2
		,
	\end{equation}
	where the fermionic operators $c_1$ and $c_2$ annihilate an electron on the
	respective dot.  The last term of $H_\text{DQD}$ accounts for inter-dot
	Coulomb interaction with strength $U$, while we assume that the repulsion
	within one dot inhibits double occupation.  We will not study spin effects
	and, thus, work with spinless electrons. The leads are described by
	$H_\text{leads} = \sum_q \epsilon_q (n_{L,q}+n_{R,q})$ with $n_{\alpha,q} =
	c_{\alpha,q}^\dagger c_{\alpha,q}$ the occupation number of mode $q$ in lead
	$\alpha=L,R$.  The coupling between the leads and the dots is given by
	\begin{equation}
		\label{HDQD}
		H_\text{DQD-leads} =
		\sum_{q} V_{L,q}c^\dagger_{L,q} c_1 %
		+\sum_{q} V_{R,q}c^\dagger_{R,q} c_2 %
		+\text{h.c.},
	\end{equation}
	where ``h.c.'' denotes the Hermitean conjugate.  The effective coupling is
	given by the spectral density $\Gamma = 2\pi\sum_q |V_{L/R,q}|^2
	\delta(\epsilon-\epsilon_q)$ which we assume energy independent.  It
	provides the dot-lead rate $\Gamma/\hbar$.
	
	Our second system is a QPC between two leads modelled by the Hamiltonian
	$H_\text{QPC} = \sum_k \epsilon_k c_k^\dagger c_k + \sum_{k'} \epsilon_{k'}
	c_{k'}^\dagger c_{k'}$, where $k$ and $k'$ label the modes of the left and
	the right lead, respectively.  The leads are weakly coupled by the
	tunnel Hamiltonian $\Lambda = \Lambda_++\Lambda_-$, where
	\begin{equation}
		\label{HQPC}
		\Lambda_+ = \sum_{k,k'} t_{kk'} c_{k'}^\dagger c_k ,
	\end{equation}
	transfers an electron from the left to the right QPC lead, while
	$\Lambda_-=\Lambda_+^\dagger$ describes the opposite process.
	In a continuum limit, the matrix elements $t_{kk'}$ are encompassed by
	the energy-independent QPC conductance $G = 2\pi \sum_{kk'}
	|t_{kk'}|^2 \delta(\epsilon-\epsilon_k) \delta(\epsilon-\epsilon_{k'})$
	in units of the conductance quantum $G_0 = e_0^2/h$.
	If an electron resides on dot~1, Coulomb repulsion enhances the barrier
	between the leads and, thus, reduces the tunnel matrix elements $t_{kk'}$.
	This effect is captured by a prefactor $x = (1-sn_1)$ in the tunnel
	Hamiltonian such that
	\begin{equation}
		\label{Hqpc}
		H_\text{QPC}^\text{tun}
		= x(\Lambda_++\Lambda_-)
		= (1-sn_1) (\Lambda_++\Lambda_-) 
	\end{equation}
	accounts for both the QPC and its coupling to the DQD.  For consistency,
	the dimensionless coupling $s$ must obey $s\leq1$.

	\subsection{Master equation and full-counting statistics}
	
	Our theoretical description is based on the formal elimination of all four
	leads such that we remain with a reduced master for the DQD.  After
	transforming the Liouville-von Neumann equation for the total density
	operator into the interaction picture with respect to $H_\text{DQD}$ and
	the lead Hamiltonians, we derive a Markovian master equation
	\cite{RedfieldIBMJRD1957a} that captures the remaining terms to second
	order,
	\begin{align}
\dot \rho = -\frac{i}{\hbar}[H_S,\rho]
-\frac{1}{\hbar^2}\sum_{n}\int\limits_{0}^{\infty} dt\,
\tr_{\textrm{leads}} [V_n, [\tilde V_{n}(-t), \rho\otimes R_0]],
\label{BRME}
	\end{align}
	where $\rho$ is the reduced DQD density operator.  $R_0$ refers to the
	grand canonical ensemble of each lead $\alpha$ with a chemical potential
	shift $\mu_\alpha$ measured with respect to a common Fermi energy
	$\epsilon_F$.  The operators $V_n$ represent $H_\text{QPC}^\text{tun}$ and
	the two tunnel contributions of $H_\text{DQD-leads}$. 
	
	While the reduced master equation \eqref{BRME} fully describes the DQD, the
	lead degrees of freedom are traced out, so that information about the
	transported electrons gets lost.  To be able to recover it, we multiply
	before tracing out the leads the full density operator $\rho\otimes R_0$ by
	a phase factor $e^{i\bm\chi\cdot\bm N}$, where the elements of the vectors
	$\bm\chi = (\chi_L, \chi_R, \chi_\text{QPC})$ and $\bm N = (N_L,N_R,
	N_\text{QPC})$ refer to the left and the right lead of the DQD and to the
	QPC, respectively.%
	\footnote{When an electron tunnels between the DQD and one of its leads,
	the DQD occupation and, hence, $N_L+N_R$ change by $\pm1$. Therefore,
	a full description requires two independent counting variables.
	For the QPC, by contrast, $N^L_\text{QPC}+N^R_\text{QPC}=\text{const}$, so
	that one counting variable is sufficient.}
	Then we obtain the master equation $\dot\rho = \mathcal L({\bm\chi})\rho$
	with the generalized Liouvillian $\mathcal L({\bm\chi})\rho
	=-(i/\hbar)[H_\text{DQD},\rho] + \mathcal L_\text{DQD-leads}({\bm\chi})\rho
	+\mathcal L_\text{QPC}({\bm\chi})\rho$.  The now $\bm\chi$-dependent
	$\rho$ obeys $\tr\rho = \langle e^{i\bm\chi\cdot\bm
	N}\rangle$, i.e., it is a generating function from which the
	moments of the lead electron distributions can be computed by taking
	derivatives with respect to components of $\bm\chi$.  Accordingly,
	$ Z(\bm\chi) \equiv \frac{\partial}{\partial t}\ln\tr\rho$ generates
	current cumulants, while derivatives at different
	times provide current correlation functions.  After multiplication with the
	appropriate power of the electron charge $-e_0$, we obtain the
	currents as the corresponding change of the lead electron number:
	$I_2 = -ie_0\partial Z/\partial\chi_\text{QPC}$ and $I_1 = (-ie_0/2)
	(\partial Z/\partial\chi_L - \partial Z/\partial\chi_R)$ evaluated at
	$\bm\chi=\bm0$.  The definition of $I_1$ is motivated by displacement
	currents in the leads which have the consequence that primarily the
	symmetrized current is experimentally accessible.  While being irrelevant
	for the average current, this affects time-dependent quantities such as
	correlation functions \cite{Blanter2000a}.
	
	The evaluation of $\mathcal{L}_\text{DQD-leads}({\bm\chi})$ for the
	incoherent DQD-lead tunneling is rather standard, see e.g.\ the appendix of
	Ref.~\cite{Hussein2012a}.  It essentially consists of jump operators
	between many-particle DQD states that differ by one electron.  The transition
	rates contain Fermi functions reflecting the initial occupation of
	the lead modes.
	The elimination of $H_\text{QPC}^\text{tun}$, by contrast, is way less
	common and, moreover, describes the interaction of the two circuits which
	is in the focus of the present article.  Therefore it is worthwhile to
	discuss it in more detail.
	
	By evaluating the $t$-integral in Eq.~\eqref{BRME} we end up with the QPC
	part master equation given by
	\begin{equation}
		\label{Lqpc(chi)}
		\mathcal{L}_\text{QPC}(\bm\chi) =
		\mathcal{L}_\text{QPC}
		+ (e^{i\chi_\text{QPC}}-1)\mathcal{J}^+
		+ (e^{-i\chi_\text{QPC}}-1)\mathcal{J}^- .
	\end{equation}
	Its first term is the QPC Liouvillian
	\begin{equation}
		\label{Lqpc0}
		\begin{split}
			\mathcal{L}_\text{QPC}\rho =
			\frac{1}{2\hbar^2}\int_{-\infty}^{+\infty} dt\; C(t)\big[ 
			& \tilde x(-t)\rho x + x\rho\tilde x(t)\\
			-& x\tilde x(-t)\rho-\rho\tilde x(t) x
			\big] .
		\end{split}
	\end{equation}
	The symmetrization of the time integral corresponds to
	neglecting energy renormalization stemming from principal values.  A main
	ingredient to $\mathcal{L}_\text{QPC}$ is the correlation function of the
	QPC tunnel operator, $C(t) = \langle \Lambda(t)\Lambda(0)\rangle =
	C_+(t)+C_-(t)$, where $C_\pm(t) =
	\langle\Lambda_\mp(t)\Lambda_\pm(0)\rangle$ is readily evaluated from its
	definition and the assumption that the leads are voltage biased.  In
	Fourier representation it reads \cite{Ingold1992a}
	\begin{align}
\label{Comega}
C_\pm(\omega)
={} & G\, \frac{\hbar\omega\pm e_0V}{1-\exp[-(\hbar\omega\pm e_0V)/k_BT]}
\\
={} & C_\mp(-\omega)e^{(\hbar\omega \pm e_0V)/k_BT}
\label{Cboltzmann}
,
	\end{align}
	see sketch in Fig.~\ref{fig:setup}(b).  For energy absorption from the
	environment, it is evaluated at negative frequencies, for emission at
	positive frequencies.  In the low-temperature limit, $C(\omega)$ vanishes
	for $\hbar\omega<-|e_0V|$, which means that the size of the energy quanta
	absorbed by the DQD is limited by the QPC bias.
	Interestingly enough, $C_\pm$ can be written in terms of $\epsilon
	[1-\exp(\epsilon/k_BT)]^{-1}$, an expression appearing in master equations
	for Ohmic quantum dissipation.
	
	The second and third term in Eq.~\eqref{Lqpc(chi)} stem from the counting
	field and vanish for $\chi_\text{QPC}\to 0$. The superoperator
	\begin{equation}
		\label{J+}
		\mathcal{J}_+\rho = \frac{1}{2\hbar^2}\int_{-\infty}^{+\infty} dt\;
		C_+(t) \big[\tilde x(-t)\rho x + x\rho\tilde x(t)\big]
	\end{equation}
	describes forward tunneling, while backward tunneling is given by the
	corresponding expression with $C_-(t)$.
	
	For the numerical treatment, we decompose the superoperators into the
	eigenstates of $H_\text{DQD}$.  In this basis, the interaction picture
	operators contain time-dependent phase factors, so that the time integrals
	yield delta functions and, finally, $C_\pm(\omega)$ has to be evaluated at
	the transition frequencies of the DQD.  Thus, we have obtained an explicit
	representation of the Liouvillian $\mathcal{L}(\bm\chi)$ which provides
	access to correlation functions \cite{Hussein2014b}, current cumulants
	\cite{Flindt2008a}, and transport coefficients \cite{Hussein2014a}.
	Details for each method can be found in the respective quoted reference.
	
	\section{Exchange fluctuation relations}
	\label{sec:xft}
	
	For typical system-lead models, one can write a formally exact expression
	for the cumulant generating function $Z$ beyond the present perturbation
	theory.  While generally such expressions cannot be evaluated exactly, they
	provide formal properties such as the so-called exchange fluctuation
	relation
	\cite{Saito2008a}
	\begin{equation}
		\label{xft}
		Z(\bm\chi) = Z(-\bm\chi-i\bm\mu/k_BT),
	\end{equation}
	where $Z(\bm\chi)$ implicitly depends on the chemical potentials $\bm\mu$.
	The derivation of this relation is based on the assumption that each lead
	consists of a continuum of modes initially in a Gibbs state, while
	the central system (here: the DQD) possesses only a few degrees of freedom.
	A main interest in such relations stems from the fact that the Taylor
	coefficients of $Z$ at $\bm\mu=\bm\chi=\bm0$ are transport coefficients,
	i.e., experimentally accessible quantities such as the conductivity.
	Hence, Eq.~\eqref{xft} provides relations between transport
	coefficients known as Onsager-Casimir relations \cite{Saito2008a}.
	
	To highlight the relation between the cumulant generating function $Z$ and
	Casimir-Onsager relations, we notice that by construction, the electron
	number in lead $\alpha$ changes by the current $I_\alpha = -i e_0\partial
	Z/\partial\chi_\alpha|_{\bm\chi=\bm0}$.  Moreover, close to equilibrium
	$\bm\mu=\bm0$, the current is linear in the voltages.  Its slope is a
	transport coefficient, namely the (trans) conductance $G_{\alpha\beta} =
	-\partial I_\alpha/\partial\mu_\beta|_{\bm\mu=\bm0}$ which obviously is a
	second derivative of $Z(\bm\chi)$.  Taking the same derivative also on the
	r.h.s.\ of Eq.~\eqref{xft} yields $S_{\alpha\beta}/k_BT - G_{\alpha\beta}$,
	where $S_{\alpha\beta}$ is the (co) variance of the currents.  Thus, we
	obtain the Johnson-Nyquist relation $2k_BT\,G_{\alpha\beta} =
	S_{\alpha\beta}$.
	Generalizing this concept, we define the transport coefficients
	\begin{equation}
		\label{K}
		K^{\alpha_1\cdots\alpha_m}_{\beta_1\cdots\beta_n}
		=
		\frac{(-i)^m\, e_0^{m+n}\,\partial^{m+n}}{\partial\chi_{\alpha_1} \cdots
		\partial\chi_{\alpha_m} \partial\mu_{\beta_1} \cdots
		\partial\mu_{\beta_n}} Z_(\bm\chi)
		\Big|_{\bm\chi=\bm\mu=\bm0} \,.
	\end{equation}
	Applying the same derivative to Eq.~\eqref{xft} provides generalized
	Casimir-Onsager relations, which we abbreviate as
	$R^{\alpha_1\cdots\alpha_m}_{\beta_1\cdots\beta_n} = 0$.  For example,
	$R_\beta^\alpha=0$ denotes the Johnson-Nyquist relation.
	
	Since approximation schemes are not necessarily consistent with exact
	formal properties, the question arises whether our master equation approach
	complies with Eq.~\eqref{xft}.  Its consistency has been verified
	for various specific situations \cite{Sanchez2010a, BulnesCuetara2011a,
	Golubev2011a, BulnesCuetara2013a, UtsumiPRB2010a, Lopez2012a}, while a
	general proof has been given only in the classical limit
	\cite{Esposito2009a, Campisi2011a} and within rotating-wave approximation
	(RWA) in the single-particle limit \cite{Esposito2009a} and for
	many-particle states \cite{Hussein2014a}.
	Beyond RWA, Eq.~\eqref{xft} may be compromised.
	The deviations, however, tend to be tiny, in particular in the
	low-temperature limit \cite{Hussein2014a} which is non-trivial since
	the temperature appears in the denominator.  Here we
	explore the situation for a master equation with the non-conventional
	Liouvillian $\mathcal{L}_\text{QPC}$.
	In doing so, we extend the results of Refs.~\cite{Utsumi2010a,
	Golubev2011a} to the presence of quantum coherence and those of
	Ref.~\cite{Hussein2014a} to the coupling to a QPC.

	\subsection{Bloch-Redfield equation in RWA}
	
	In the limit of very weak coupling between the central system and the
	leads, the reduced density operator becomes eventually diagonal in the
	energy basis defined by $H_\text{DQD}|\alpha\rangle =
	E_\alpha|\alpha\rangle$.  Thus, one often can work with the ansatz
	$\rho_{\alpha\beta} = P_\alpha\delta_{\alpha\beta}$ which provides the
	Pauli-type master equation $\dot P_\alpha = \sum_{\alpha'}
	W_{\alpha\alpha'}(\bm\chi) P_{\alpha'}$.  The transitions rates
	$W_{\alpha\alpha'}(\bm\chi)$ follow from a basis decomposition of the
	master equation~\eqref{BRME} after having introduced the counting variables
	$\bm\chi = (\chi_L,\chi_R,\chi_\text{QPC})$.  For our later reasoning, it
	is important that the long-time solution of a Markovian master equation is
	dominated by the eigenvalue $-\lambda_0(\bm\chi)$ of the corresponding
	matrix $W(\bm\chi)$ that vanishes in the limit $\bm\chi\to\bm0$, i.e., the
	one that corresponds to the stationary solution in the absence of the
	counting variable.  Then $P\sim e^{-\lambda_0(\bm\chi)t}$ which translates
	to $Z=-\lambda_0$ and implies that the generating function can be
	traced back to the computation of an eigenvalue of $W(\bm\chi)$
	\cite{Bagrets2003a}.  Hence we can conclude that our RWA master equation
	complies with Eq.~\eqref{xft} if $W(\bm\chi)$ and
	$W(-\bm\chi-i\bm\mu/k_BT)$ are isospectral.  For the Liouvillian of the
	type $\mathcal{L}_\text{DQD-leads}$ for the dot-lead tunneling,
	this has already been demonstrated in Ref.~\cite{Hussein2014a}.  Thus, we
	can restrict ourselves to the corresponding calculation for
	$\mathcal{L}_\text{QPC}(\bm\chi)$ in Eq.~\eqref{Lqpc(chi)}.
	
	By evaluating $W_{\alpha\alpha'}(\bm\chi)$ for $\mathcal{L}_\text{QPC}$ and
	$\mathcal{J}_\pm$, see Eqs.~\eqref{Lqpc0} and \eqref{J+}, we find
	\begin{equation}
		W_{\alpha\alpha'}(\bm\chi)
		= e^{i\chi_\text{QPC}} w_{\alpha\leftarrow\alpha'}^+
		+ e^{-i\chi_\text{QPC}} w_{\alpha\leftarrow\alpha'}^-
		, \quad\alpha\neq\alpha'
	\end{equation}
	with the QPC forward and backward rates
	\begin{equation}
		w_{\alpha\leftarrow\alpha'}^\pm = C_\pm(E_{\alpha'}-E_\alpha)
		|\langle\alpha|x|\alpha'\rangle|^2 .
	\end{equation}
	Since the Liouvillian $\mathcal{L}(\bm0)$ conserves the trace of
	the density matrix, we can obtain the diagonal elements $W_{\alpha\alpha}$
	from the normalization condition $\sum_\alpha \dot P_\alpha=0$ yielding
	$W_{\alpha\alpha} = -\sum_{\alpha'\neq\alpha} W_{\alpha'\alpha}(\bm0)$.
	Using Eq.~\eqref{Cboltzmann} for $\mu = -e_0V$, we find for the QPC
	contribution by straightforward algebra the relation
	\begin{equation}
		W_{\alpha\alpha'}(-\chi_\text{QPC}-i\mu/k_BT)
		= e^{E_{\alpha'}/k_BT} W_{\alpha'\alpha}(\chi_\text{QPC}) e^{-E_\alpha/k_BT}.
	\end{equation}
	Together with the corresponding relation for the dot-lead tunneling
	\cite{Hussein2014a}, we obtain
	\begin{equation}
		W(\bm\chi)^T = S^{-1}W(-\bm\chi-i\bm\mu/k_BT)S, \quad
		S =  e^{-H_\text{DQD}/k_BT},
	\end{equation}
	where $T$ denotes matrix transposition.  This means that the RWA
	Liouvillian with the modified counting variable relates to the original one
	by a transposition and a similarity transformation which both leave the
	eigenvalues invariant.  Therefore, we can conclude that the RWA
	master equation agrees with the exchange fluctuation relation~\eqref{xft}.
	
	\subsection{Deviations and experimental tests}
	
	Despite possessing the desirable consistency with Eq.~\eqref{xft},
	a RWA master equation has obvious limitations when off-diagonal
	density matrix elements play a role.  This is, e.g., the case for a strongly
	biased, but undetuned DQD in which the inter-dot tunneling is weaker than
	the dot-lead tunneling.  Then an electron that enters from the source
	will stay on dot~1 for a while until it proceeds to dot~2.  From
	there it will tunnel rapidly to the drain.  Thus, the DQD
	will will predominantly be in the state $c_1^\dagger|0\rangle$, so that its
	density operator in the energy basis has a large off-diagonal contribution
	and the RWA will fail \cite{Kaiser2006a}.  As a drawback, however, a treatment
	beyond RWA is generally not fully consistent with exchange fluctuation
	relations \cite{Hussein2014a}, which motivates a quantitative study of
	possible discrepancies.
	
	\begin{figure}[t]
		\includegraphics{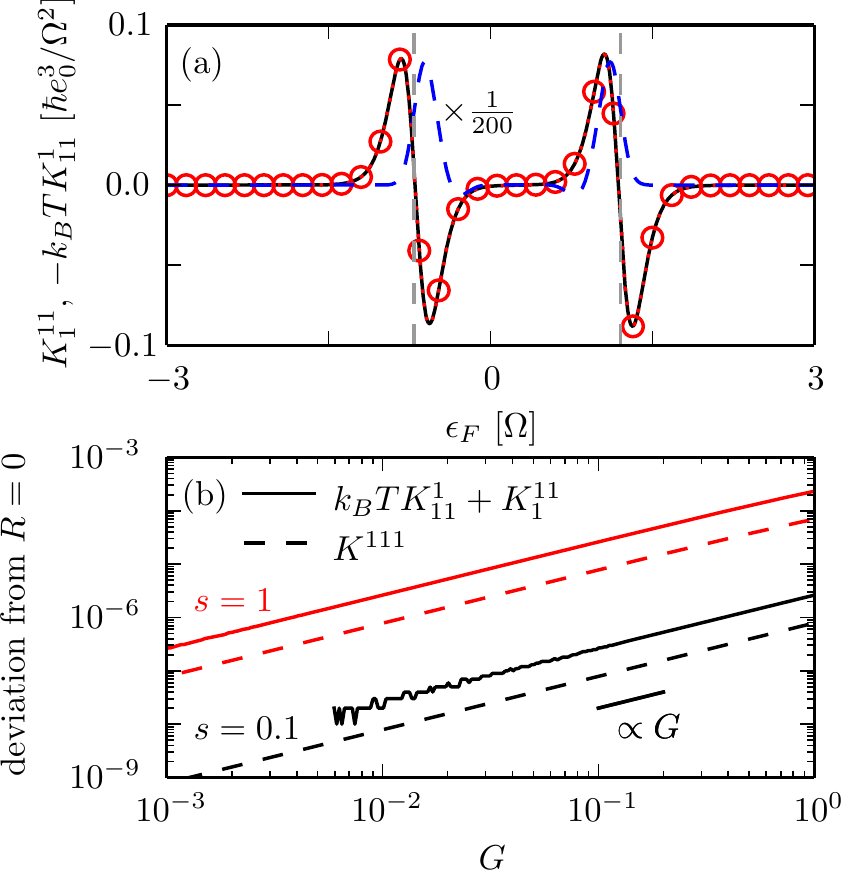}\caption{\label{fig:cft}
		(a) Transport coefficients $K_1^{11}$ (solid line) and $-K_{11}^1$
		(circles) computed with our Bloch-Redfield master equation as a function of
		the Fermi energy for coupling $s=1$, QPC conductance $G=1$,
		temperature $T=0.1\Omega/k_B$, and the DQD parameters
		$\Gamma=\epsilon=2U=\Omega$.  While according to the exchange
		fluctuation relation must be differ only by a factor $k_BT$, the master
		equation predicts an additional tiny spurious deviation (dashed line).  The
		vertical lines mark the location of the conductance peaks as discussed in
		the text.
		(b) Scaling of the deviations for $k_BT K_{11}^1+K_1^{11}$ and
		$K^{111}$ from their exact theoretical value zero as a
		function of the QPC conductance for $\epsilon_F=U+\Delta/2$.
		}
	\end{figure}%
	Experimentally, the cumulant generating function is not directly accessible
	and, thus, one may test instead Casimir-Onsager relations derived from it.
	We here focus on transport coefficients that have been tested with an
	Aharonov-Bohm ring \cite{Nakamura2010a} and a DQD coupled to a QPC
	\cite{Utsumi2010a}.
	We start with the Johnson-Nyquist relation and find in accordance with
	Ref.~\cite{Hussein2014a} that it is fulfilled within our numerical
	precision.
	
	Thus, for revealing possible discrepancies, we have to employ
	Casimir-Onsager relations of higher order such as $k_BTK_{11}^1+K_1^{11} =
	0$ which follows directly from $R_{11}^1=0$ and $R^{111}=0$.  Figure
	\ref{fig:cft}(a) depicts the master equation result for both terms involved
	in this relation.  Their main contribution appears at the conductance
	peaks, i.e., at Fermi energies that obey $E^{(n+1)} = E^{(n)}+\epsilon_F$,
	where $E^{(n)}$ is the energy of the lowest state with $n$ electrons on the
	DQD.  Here, $E^{(0)}=0$, $E^{(1)}=-\frac{1}{2}\sqrt{\epsilon^2+\Omega^2}$,
	and $E^{(2)}=U$, as can be readily verified by diagonalizing $H_\text{DQD}$
	in Fock basis. The interpretation of this condition is that an electron
	from the Fermi surface can tunnel between the lead and the DQD without
	energy cost, so that already a small voltage can induce a current. 
	With the scale chosen, $k_BTK_{11}^1$ and $K_1^{11}$ appear identical as
	they ideally should be.  A closer look, however, reveals a relative
	difference of the order $10^{-3}$.  Such tiny differences are beyond the
	resolution of related experiments \cite{Nakamura2010a, Utsumi2010a}, which
	lets us conclude that our master equation treatment is sufficiently precise
	to compute the quantities employed for typical tests of exchange fluctuation
	relations.
	
	Beyond such practical issues, it is interesting to see how these
	discrepancies scale with the system parameters.  For the ``conventional''
	dot-lead jump operators such as those in $\mathcal{L}_\text{DQD-leads}$, one
	generically finds for small dot-lead rate $R \propto \Gamma^3$, while for
	small inter-dot tunneling, $R\propto\Omega^2$ \cite{Hussein2014a}.  For
	$\mathcal{L}_\text{QPC}$, see Eq.~\eqref{Lqpc(chi)}, the relevant
	parameters are the dimensionless coupling $s$ and the QPC conductance $G$,
	which brings the scaling as a function of $s$ and $G$ to our attention.
	Figure~\ref{fig:cft}(b) depicts the peak values of the deviation shown in
	Fig.~\ref{fig:cft}(a). It indicates that $k_BTK_{11}^1+K_1^{11}\propto
	s^2G$.  In particular for $s=0.1$, the l.h.s.\ of this relation involves
	computing the tiny difference between two much larger numbers, which is at
	the limit of our numerical precision.  Indeed one notices that for small
	values of $G$, the data is compromised by numerical errors.  We therefore
	also studied the Casimir-Onsager relation $K^{111}=0$ which follows from
	$R^{111}=0$.  Since this relation consists of only one transport
	coefficient, it does not suffer from the mentioned numerical problem.  The
	corresponding data in Fig.~\ref{fig:cft}(b) confirms for the deviations the
	scaling $\propto G$ and $\propto s^2$.
	
	Thus we can conclude that the full Bloch-Redfield master equation for the
	QPC coupling is consistent with exchange fluctuation relations up to
	corrections $R\propto s^2G$, i.e, corrections linear in $G$.  By
	contrast, the mentioned scaling with the DQD parameters $\Gamma$ and
	$\Omega$ is more favorable \cite{Hussein2014a}.  Nevertheless, the
	Bloch-Redfield approach can be safely applied for the computation of
	transport coefficients up to second order and when a RWA treatment is
	sufficient.
	
	\section{Tunnel contact as driving source}
	\label{sec:ratchet}
	
	When non-equilibrium noise acts upon the DQD, it will induce electron
	transitions from the ground state of the DQD to the excited state, see
	Fig.~\ref{fig:setup}(a).  Once in the excited state, the electron will
	leave predominantly to the right lead, while subsequently an electron from
	the left lead enters.  In this way, the QPC can spurs a directed current in
	an unbiased DQD circuit.  Its direction depends on the sign of $\epsilon$,
	which implies a current reversal point at $\epsilon=0$.  For strong DQD
	detuning, this mechanism has been proposed as noise detector
	\cite{Aguado2000a}.  Going beyond that work, our master equation
	description starts from a complete model that includes the noise source.
	Moreover, it allows us to study also the backaction on the QPC as well as
	cross correlations between the circuits.
	
	\subsection{Ratchet current}
	
	To estimate the ratchet current, we compute the rates for the scenario
	sketched above.  A main role is played by electron transitions from the
	one-electron ground state of the DQD, $|g\rangle = |1\rangle\cos\phi +
	|2\rangle\sin\phi$ to the excited state $|e\rangle = -|1\rangle\sin\phi +
	|2\rangle\cos\phi$, where $\cos(2\phi) = \epsilon/\Delta$ and $|n\rangle =
	c_n^\dagger|0\rangle$.  These transitions are induced by the
	Hamiltonian~\eqref{Hqpc}, and occur with the golden-rule rate $\gamma
	= (s^2/4) \sin^2(2\phi) C(-\Delta/\hbar)/\hbar$.  The first factor of this
	expression stems from the matrix element $\langle e|n_1|g \rangle$ and
	accounts for the delocalization of the DQD orbitals.
	
	When the electron has reached the excited state, it can leave the DQD
	through either lead, with a probability that depends on the overlap of
	$|e\rangle$ with the localized states $|1\rangle$ and $|2\rangle$.  The
	corresponding probabilities read $|\langle 1|e\rangle|^2 = \sin^2\phi$ and
	$|\langle 2|e\rangle|^2\cos^2\phi$.  Since these transitions contribute to
	the current with opposite sign, we finally obtain $I =
	e_0\gamma(\cos^2\phi-\sin^2\phi)$, which can be written as
	\begin{equation}
		\label{Ir}
		I
		= \frac{e_0}{\hbar} \frac{s^2 \epsilon\Omega^2}{4(\epsilon^2+\Omega^2)^{3/2}}
		C(-\Delta/\hbar).
	\end{equation}
	This expression generalizes the result of Ref.~\cite{Aguado2000a} to the
	regime $|\epsilon| \lesssim\Omega$, in which we expect current reversals as
	a most relevant feature for possible applications.
	
	\begin{figure}[t]
		\centering
		\includegraphics{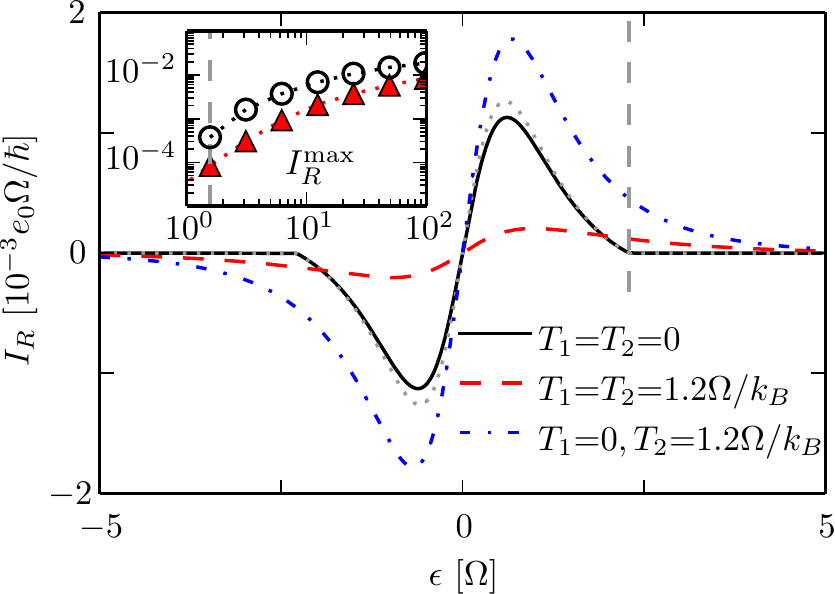}\caption{\label{fig.:ratchetCurrent}
		Ratchet current through the DQD as a function of the detuning $\epsilon$
		for tunneling matrix element $\Omega=10\Gamma$, QPC conductance
		$G=1$ and voltage $V=2.5\Omega/e_0$, coupling strength $s=0.1$,
		interaction $U=50\Omega$, and
		various temperatures.  The dotted line marks the analytical solution
		\eqref{Ir} for $k_BT=0$.  The vertical lines marks the condition
		$\epsilon^2 +\Omega^2 =(e_0V)^2$ beyond which the DQD splitting is always
		larger than $e_0V$.
		Inset: current maximum as a function of the QPC bias voltage $V$ for
		$k_BT=0$ (circles) and $k_BT=1.2\Omega$ (triangles) in units of
		$e_0\Omega/\hbar$.
		}
	\end{figure}%
	
	Our numerical solution allows refining the picture drawn by the golden-rule
	calculation.  Figure~\ref{fig.:ratchetCurrent} shows the ratchet current as
	a function of the DQD detuning, which for zero temperature by and large
	confirms the behavior predicted by Eq.~\eqref{Ir}, but indicates that the
	analytic approach overestimates the ratchet effect.  When the
	QPC temperature is increased, its current acquires a thermal component so
	that the effective driving becomes stronger.  Accordingly, we witness an
	enhancement of the the ratchet current.  By contrast, when both the QPC
	temperature and the DQD temperature are increased simultaneously, the bias
	voltage plays a smaller role and both circuits are dominated by thermal
	noise.  Eventually the system reaches thermal equilibrium in which the
	current vanishes.  The dashed line in Fig.~\ref{fig.:ratchetCurrent} shows
	that already for the moderate temperature $T=1.2 \Omega/k_B$, the ratchet
	current is significantly smaller than at $T=0$.
	
	The limitations of our analytic approach are also visible in the current
	maxima.  While Eq.~\eqref{Ir} for $e_0V\gg\epsilon$ predicts a linear
	increase with $V$, the data shown in the inset of
	Fig.~\ref{fig.:ratchetCurrent} demonstrates a sub-linear growth.  This
	deviation is even more clearly visible in the ratchet current as a function
	of the QPC bias $V$ for constant detuning (inset of
	Fig.~\ref{fig.:FanoFactor}).  The main reason for this discrepancy is that
	our analytic approach is based on the assumption that the intra-dot
	excitation is the slowest process and, thus, governs the dynamics.  Upon
	increasing $V$, however, this assumption will be violated at some stage and
	we would have to employ rate equation description along the lines of
	Ref.~\cite{Stark2010a} which considers also the de-excitation
	$|e\rangle\to|g\rangle$.

	\subsection{Current noise and cross correlations}
	
	\begin{figure}[t]
		\centering
		\includegraphics{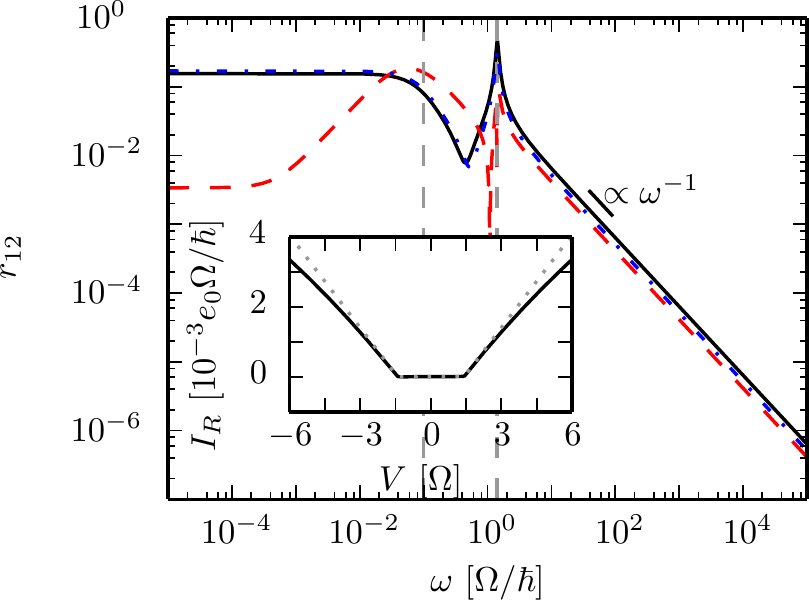}\caption{\label{fig.:FanoFactor}
		Absolute value of the correlation coefficient $r_{12}$ between the DQD current and the QPC
		current as a function of the frequency for the detuning $\epsilon=\Omega$.
		All other parameters are as in Fig.~\ref{fig.:ratchetCurrent}.  The
		vertical lines mark the values $\omega=\Gamma/\hbar$ and
		$\omega=\Delta/\hbar$, respectively.  The inset shows the ratchet current
		in dependence on the applied bias voltage for zero temperature (solid line)
		and for $k_B T=1.2\Omega$ (dotted).}
	\end{figure}%
	
	To characterize current fluctuations, we consider the correlation functions
	$S_{\alpha\beta}(t-t') = \langle I_\alpha(t)I_\beta(t')- \langle
	I_\alpha\rangle\langle I_\beta\rangle$, where $\alpha$ and $\beta$ label
	the two circuits.  For the auto-correlations we find in the frequency domain
	roughly $S_{\alpha\alpha}(\omega) \approx \langle I_\alpha\rangle$ (not
	shown), i.e., the typical frequency-independent value for the shot noise of
	a Poisson process \cite{vanKampen1992a}.  The physical picture behind this
	behavior is that of uncorrelated tunnel events which holds for a tunnel
	contact in the low-temperature regime as well as for the DQD in the regime
	in which the transitions $|g\rangle\to|e\rangle$ represent the bottleneck
	for the electron flow.
	
	Let us turn to the cross correlation $S_{12}$ which contains information
	about the mutual influence of the circuits on each other.  A qualitative
	picture of the physical process emerges from normalized correlations.
	In the spirit of full-counting statistics of mesoscopic transport, it
	is common to define the Fano factors $F_{\alpha\alpha}(\omega) =
	S_{\alpha\alpha}(\omega)/I_\alpha$ and the cross Fano factor
	$F_{12}(\omega) = S_{12}(\omega)/\sqrt{I_1 I_2}$.  These quantities relate
	to the $g^{(2)}$ function in optics and hint on bunching
	and anti-bunching of transport events \cite{Blanter2000a, Emary2012a}.
	In a typical experimental realization, the currents in the two circuits may
	differ by several orders of magnitude \cite{Khrapai2006a}. Therefore we are
	not so much interested in correlated counting, but rather in the correlation
	coefficient $r_{12}(\omega) = S_{12}(\omega)/\sqrt{S_{11}(\omega)
	S_{22}(\omega)}$ which is normalized to the frequency-dependent
	auto-correlations.  In our case, $r_{12}(\omega)$ and
	$F_{12}(\omega)$ are practically the same owing to the prevailing
	Poissonian nature of each current and the corresponding almost constant
	$S_{\alpha\alpha}(\omega)$.
	
	Figure~\ref{fig.:FanoFactor} shows the correlation coefficient for various
	temperatures.  It is characterized by two energies.  Below the dot-lead
	coupling $\Gamma$, we find a plateau after which a decay
	$\propto\omega^{-1}$ sets in.  At the level splitting of the DQD, $\Delta$,
	we find a peak with a value up to $r_{12}(\Delta/\hbar)\approx 0.85$
	indicating strong correlations between the ratchet current and the QPC at
	resonance.  Thus, correlations mainly appear when the QPC can induce
	resonant transitions between DQD orbitals.  The sharp Lorentzian form of
	the peak indicates the relevance of quantum coherence, which implies that
	this feature is beyond the treatment with rate-equations that leads to the
	analytic solution \eqref{Ir}.  For smaller values of $\Delta$, coherence
	eventually gets lost and the peak submerges below the low-frequency
	plateau.  As expected for a coherence effect, the peak becomes smaller at
	higher temperatures.

	\section{Conclusions}
	
	We have investigated a setup composed of two capacitively
	interacting nano conductors, namely a QPC in the weak tunneling limit and a
	DQD.  While in most recent experimental realization of such setups, the QPC
	serves as charge monitor for the DQD, we were interested in aspects beyond
	that.  In particular, we focused on the non-equilibrium DQD
	dynamics induced by the shot noise of the QPC current and on testing the
	validity of exchange fluctuation relations.
	
	For the theoretical treatment, we have derived a Bloch-Redfield master
	equation augmented by a counting variable for the electron number in each
	lead.  This allows one to eliminate the leads within second-order
	perturbation theory without losing information about the transported
	electrons.  For the dot-lead tunneling, one obtains ``conventional'' jump
	operators, while the coupling to the QPC yields dissipative terms
	resembling those stemming from the coupling to a bosonic heat bath.
	We demonstrated that within RWA, the resulting unconventional jump
	operators are fully consistent with exchange fluctuation relations.
	Beyond RWA, we numerically found deviations which decay linearly
	with the QPC conductance and quadratically with the coupling.
	In the experimentally relevant regime, these deviations are rather small and
	do not inhibit computing with a Bloch-Redfield formalism transport
	coefficients for testing Casimir-Onsager relations.  The Johnson-Nyquist
	relation even turned out to be fulfilled within our numerical precision.
	
	When using the QPC as a driving source, one can induce a ratchet current in
	the DQD.  As a feature most relevant for applications, it possesses a
	current reversal as a function of the detuning.  While generally the
	ratchet current reflects the spectrum of the effective noise that the QPC
	entails on the DQD, our numerical results revealed significant deviations
	from the behavior found with a simple golden-rule treatment.  The
	correlations between the currents in both circuits are most pronounced at
	frequencies that correspond to the energy splitting of the DQD, which
	emphasizes the role of quantum coherence.
	For a typical inter-dot tunnel coupling of $\Omega=25\mu\mathrm{eV}$, the
	ratchet current is of the order of several $\mathrm{pA}$, a value that can
	be measured straightforwardly with present techniques.
	Given that the setup investigated is common in today's quantum dot design,
	we are convinced that our results will inspire future experiments.
	
	\begin{acknowledgements}
		This work was supported by the Spanish Ministry of Economy and
		Competitiveness via grant No.\ MAT2011-24331.
	\end{acknowledgements}
	
	\bibliographystyle{andp2012}
	\providecommand{\WileyBibTextsc}{}
	\let\textsc\WileyBibTextsc
	\providecommand{\othercit}{}
	\providecommand{\jr}[1]{#1}
	\providecommand{\etal}{~et~al.}

\end{document}